\documentclass[iop]{emulateapj}

\usepackage{apjfonts}
\usepackage{epsfig}
\usepackage{epstopdf}
\usepackage{natbib}
\usepackage{amssymb,latexsym}
\usepackage{graphicx}
\usepackage{appendix}
\usepackage{amsmath}

\newcommand{\be}{\begin{equation}}
\newcommand{\ee}{\end{equation}}
\newcommand{\bee}{\begin{eqnarray}}
\newcommand{\eee}{\end{eqnarray}}

\newcommand{\om}{ \Omega_{\rm m} }

\newcommand{\mvir}{\hbox{$M_{\rm vir}$}}

\newcommand{\vmax}{{\rm v}_{\rm max}}

\newcommand{\sigate}{ \sigma_{8}}

\newcommand{\Msun}{{\rm M}_\odot}

\newcommand{\mpc}{\rm{Mpc}}

\newcommand{\lsim}{\lower.5ex\hbox{\ltsima}}
\newcommand{\gsim}{\lower.5ex\hbox{\gtsima}}
\newcommand{\ltsima}{$\; \buildrel < \over \sim \;$}
\newcommand{\gtsima}{$\; \buildrel > \over \sim \;$}

\newcommand{\comments}[1]{}

\newcommand{\vpeak}{\rm{v}_{\rm{peak}}}
\newcommand{\vnow}{\rm{v}_{\rm{max}}}

\newcommand{\lstar}{L^*}
\newcommand{\nrich}{N_{200}}

\usepackage[usenames]{color}

\shortauthors{Reddick et al}
\shorttitle{Cosmological Constraints with Arbitrary Halo Occupation}

\begin{document}
\title{Cosmological Constraints from Galaxy Clustering and the Mass-to-Number Ratio of Galaxy Clusters: 
Marginalizing over the Physics of Galaxy Formation}
\author{Rachel Reddick\altaffilmark{1}, Jeremy L. Tinker\altaffilmark{2},
  Risa H. Wechsler\altaffilmark{1}, Yu Lu\altaffilmark{1}}
\affil{$^{1}$Kavli Institute for Particle Astrophysics and Cosmology \\
 Physics Department, Stanford University, Stanford, CA, 94305\\
 SLAC National Accelerator Laboratory, Menlo Park, CA, 94025\\
 rmredd, rwechsler@stanford.edu\\
$^{2}$ Center for Cosmology and Particle Physics, Department of Physics, New York University
}

\begin{abstract} Many approaches to obtaining cosmological constraints rely on the connection between galaxies and dark matter.  However, the distribution of galaxies is dependent on their formation and evolution as well as the cosmological model, and galaxy formation is still not a well-constrained process.  Thus, methods that probe cosmology using galaxies as a tracer for dark matter must be able to accurately estimate the cosmological parameters without knowing the details of galaxy formation a priori.  We apply this reasoning to the method of obtaining $\om$ and $\sigate$ from galaxy clustering combined with the mass-to-number ratio of galaxy clusters.  To test the sensitivity of this method to variations due to galaxy formation, we consider several different models applied to the same cosmological dark matter simulation.  The cosmological parameters are then estimated using the observables in each model, marginalizing over the parameters of the Halo Occupation Distribution (HOD).  We find that for models where the galaxies can be well represented by a parameterized HOD, this method can successfully extract the desired cosmological parameters for a wide range of galaxy formation prescriptions.  \end{abstract}

\keywords{cosmological parameters ---  large-scale structure of universe --- galaxies: clusters: general --- galaxies: halos --- galaxies: formation}

\maketitle

\section{Introduction}

Galaxies trace the dark matter distribution in the Universe in a biased and non-trivial way which depends on the properties of the galaxy sample.  However, the connection between galaxies and dark matter halos makes determining their bias feasible.  Dark matter halos are collapsed, virialized structures, which grow from from initial density perturbations in the early universe.  Galaxies form at the centers of the potential wells of these halos.  Therefore, the distribution of galaxies is closely related to the dark matter distribution in the universe.  The bias of a galaxy population depends on (1) the cosmology, which determines the statistics of the halo population, and (2) the processes of galaxy formation going on within each halo.  This galaxy bias must be accounted for when using measures such as galaxy clustering to infer properties of the dark matter distribution and the underlying cosmology.  In particular, differences in predictions for galaxy formation from that of actual galaxies may be mistaken for differences instead in the dark matter and cosmology.

Previous studies have focused on using galaxy clustering alone to determine the cosmology or 
a parameterized relationship between galaxies and dark matter.  A common approach is to use a halo occupation distribution (HOD) to specify the relationship between galaxies and dark matter halos, and then to constrain the parameters of this relationship using measurements such as galaxy two-point clustering (e.g., \citealt{YMB2008,YMB2009, Zeh2011, LTB2012} and references therein).  Other methods of inferring the relationship between galaxies and dark matter include satellite kinematics \citep{More2009}, galaxy group catalogs \citep{YMB2009}, abundance matching \citep[][and references therein]{Red2013} and counts-in-cells \citep{Reid2009}.

In parallel to these methods and to take advantage of upcoming galaxy cluster results, here we examine the method described in \cite{Tin2012}.  This approach combines a two-point clustering measurement with the mass-to-number (M/N) cluster statistic.  M/N is the ratio between the mass of the cluster determined via weak lensing and the number of galaxies it contains of some well-defined type, typically specified by cuts in luminosity and color.  The clustering is measured on all galaxies in the sample, while M/N isolates only the galaxies in the most highly biased halos. The information in M/N is analogous to that in the mass-to-light (M/L) ratio of clusters, which was combined with halo occupation techniques by \cite{vdB2003} and \cite{Tin2005} to break the degeneracy between cosmology and galaxy bias.  In this paper, we will demonstrate that this approach accurately recovers the cosmological parameters $\om$ and $\sigate$ for a variety of different galaxy formation models.  These models include significant differences in the efficiency of galaxy formation, producing different galaxy populations over the same sample of halos.  In sum, we show that including the M/N statistic with galaxy clustering allows us to marginalize over galaxy formation and galaxy bias to accurately measure cosmological parameters.

Ongoing surveys, such as the Baryon Oscillation Spectroscopic Survey (BOSS) \citep{Daw2013} and the Dark Energy Survey \citep{DES2006} provide, or will provide, galaxies over a large enough volume and depth to make precise statements about their properties.  In relation to these large surveys, there have been significant recent improvements in optical galaxy cluster finding (e.g., \citealt{Ryk2012}).  The improvements include accurate cluster photometric redshifts and richness estimates, both of which are important to accurately determining the dark matter distribution.

We will begin by discussing the different galaxy models we use in \S\ref{sec:model}.  The M/N method will be briefly reviewed in \S\ref{sec:method}.  In \S\ref{sec:results} we will present the results of applying this method to our different models.  We will summarize our results and discuss future directions in \S\ref{sec:conclude}.
Unless stated otherwise, all values are given for h$=1$, where $\rm{h}=H_0/(100~\rm{km/s}/\mpc)$.  

\section{Models}\label{sec:model}

\begin{figure}
\vspace{-0.3cm}
\includegraphics[width=0.43\textwidth]{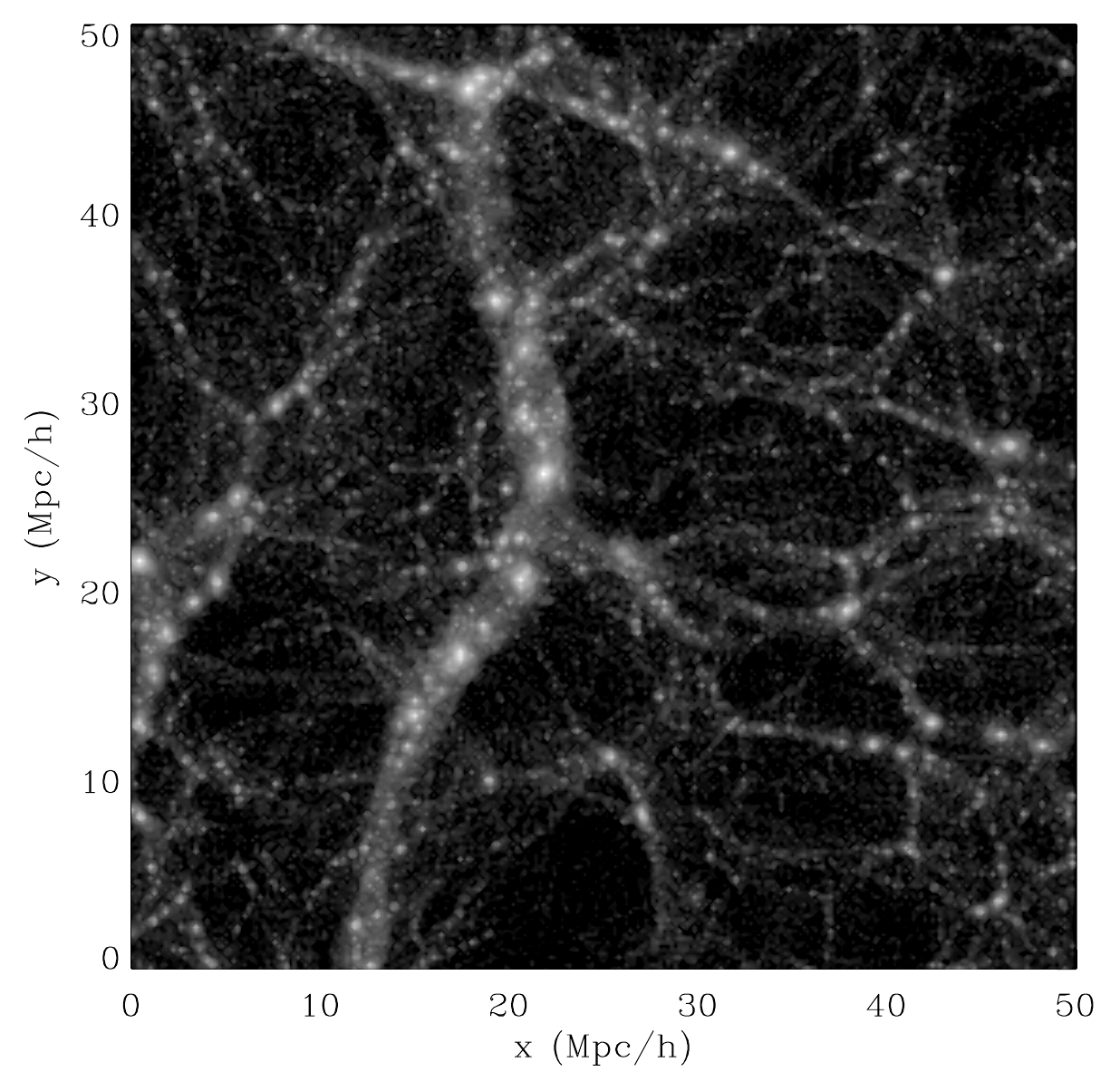}
\vspace{-0.3cm}\\
\includegraphics[width=0.43\textwidth]{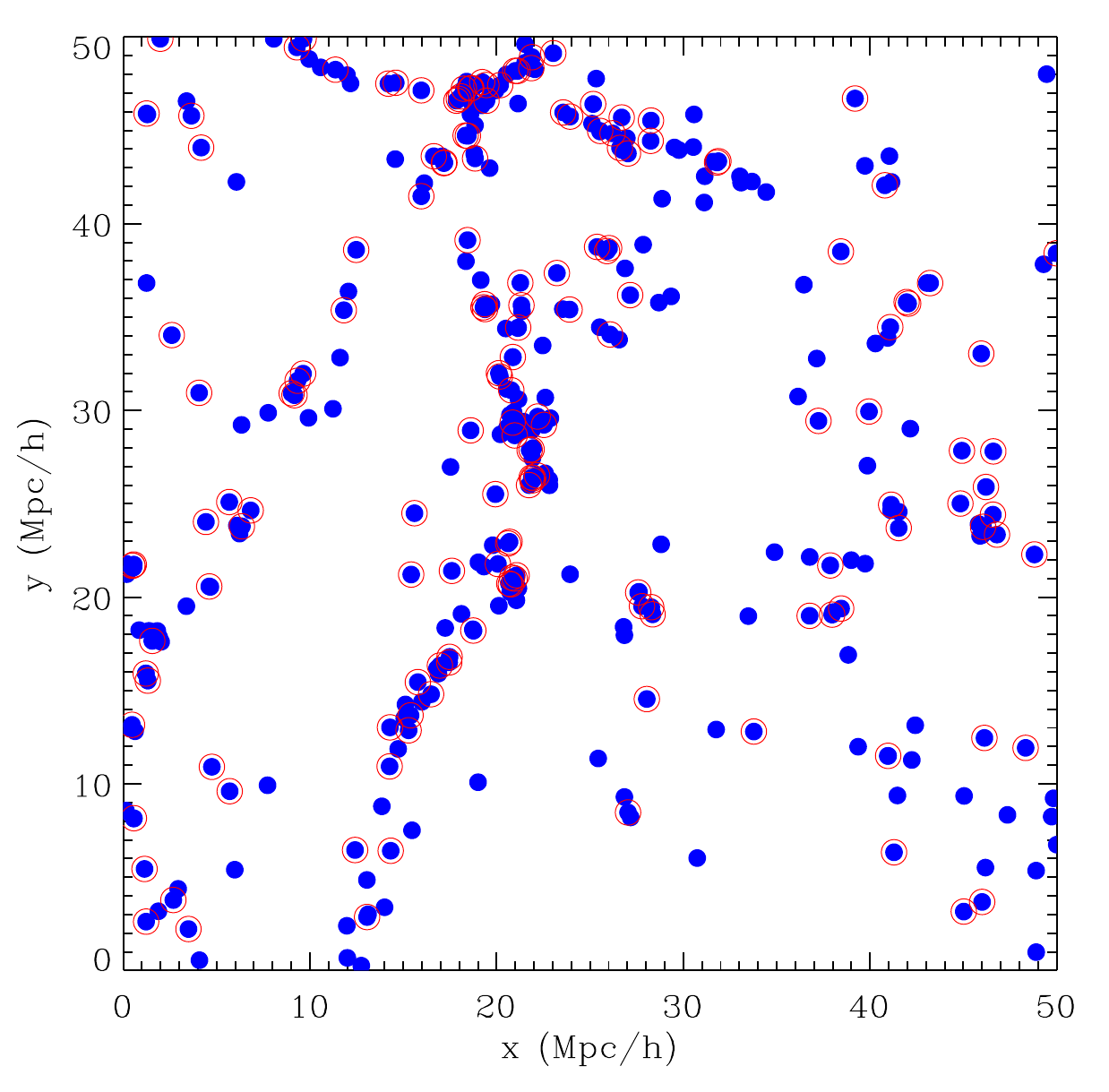}
\vspace{-0.3cm}\\
\includegraphics[width=0.43\textwidth]{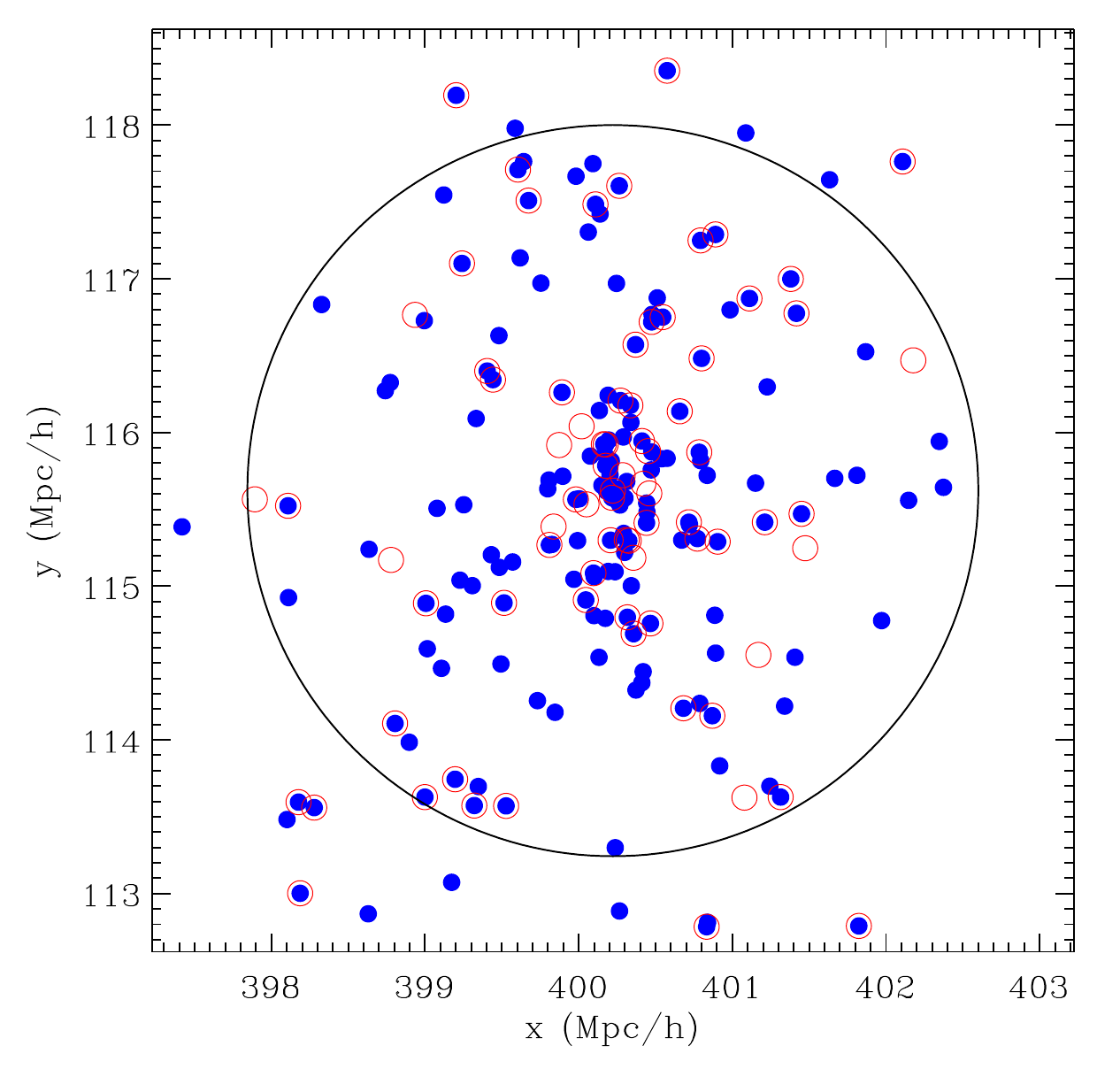}
\vspace{-0.25cm}
\caption{Visual demonstration of the difference between the two SAMs.  {\em Top:}  Dark matter distribution for a slice with thickness of 10 $\mpc/h$ in the z direction.  {\em Center:}  Open red circles indicate galaxies from the fiducial v1 SAM with $M_r-5 \log(h)<-20.5$.  Solid blue circles mark galaxies with the same cut in the v2 SAM.  The slice has a depth of 10 Mpc/h in the z direction.  {\em Bottom:}  Zoom-in on the most massive halo, with $\mvir~1e15$.  Galaxy symbols are the same; the black line indicates the virial radius.}
\label{fig:spatial}
\end{figure}

We use a handful of cosmological simulation boxes with several galaxy formation models applied.  First, we use the Consuelo box from the LasDamas suite of N-body simulations \footnote{The LasDamas simulations are available at http://lss.phy.vanderbilt.edu/lasdamas/}.  The underlying cosmology uses $\om=0.25$ and $\sigate=0.8$.  The \textsc{rockstar} halofinder \citep{BWW2013} is used for halo identification along with merger trees drawn from \cite{BWW2013b}.  The mass resolution of $\sim2\times10^9$ allows us to resolve halos that typically host galaxies below $\lstar$.  To this simulation, we apply two different methods of generating galaxy models.  The first method is a semi-analytic model (SAM) based on the methods of \cite{Lu2011} and \cite{Lu2012}, and the second is an abundance matching
method as described in \cite{Red2013}.

For the first of the two semi-analytic models, we adopt a SAM that includes flexible parameterizations for the most important baryonic processes in galaxy formation to follow the evolution of galaxies in a set of dark matter halo merger trees. The baryonic processes implemented in this SAM are described in \cite{Lu2011,Lu2012}.  Different from the previously published versions, the version of the model adopted for this paper is applied on a set of halo merger trees extracted from the Consuelo simulation, which is a cosmological N-body simulation with subhalos identified.  We consider two different versions of this SAM. One version adopts a set of parameters randomly chosen from the posterior distribution of the model that was constrained by the K-band luminosity function of local galaxies \citep{Lu2012}. Therefore, the model reproduces the K-band luminosity function of low-redshift galaxies.  We refer this version as SAM v1.  We note that because the model was constrained on Monte Carlo merger trees generated based on extended Press-Schechter (EPS) theory, when the SAM is applied to the simulation merger trees, the model reproduces the constraining data with a larger discrepancy.

The second semi-analytic model is unconstrained by observational data, and called SAM v2.  We arbitrarily change two parameters from SAM v1.  Specifically, we increase the mass scale for the halo quenching by one order of magnitude and decrease the supernova feedback mass-loading factor by a factor of 2.  These modifications result in an enhanced cooling of hot gas in high-mass halos and a reduced star formation feedback in low-mass halos.  The intent is to compare the results from a model with a physically motivated connection between galaxies and dark matter halos (SAM v1) with a model deliberately chosen to differ greatly from physical realism (SAM v2). Thus, the way galaxies form is very different between the two models, producing a dramatically different galaxy--halo connection.  The observational measures we use to infer cosmology are substantially different between the two models, even though the halo population is the same.  For instance, SAM v2 has many more bright galaxies than SAM v1.  A comparison of the spatial distribution of the galaxies in the two SAMs is shown in Fig.~\ref{fig:spatial}.

The second method matches galaxies to halos and subhalos using the abundance matching techniques of \cite{Red2013}.  This method explicitly matches the observed luminosity function from the Sloan Digital Sky Survey, and associates these luminosities with the mass, or a mass proxy, of dark matter halos and subhalos in the simulation.  The brightest galaxy expected in the volume is assigned to the most massive halo (or with the highest mass proxy), second brightest to second most massive, and so on.  For our mass proxy, we use the peak maximum circular velocity that each halo (or subhalo) has had over its history.  Some degree of scatter in luminosity at fixed "mass" is also included.  This will be referred to as the SHAM (subhalo abundance matching) model.  We use five different models.  Four are based on matching to $\vpeak$, the peak maximum circular velocity a halo or subhalo had over its history, stepping evenly in scatter from 0 to 0.3 dex in luminosity at fixed $\vpeak$.  The remaining model using the maximum circular velocity at the present time, $\vnow$, with 0.2 dex scatter.  Each of these models has a different relationship between galaxies and dark matter, which changes the halo occupation even with an unchanged luminosity function.  These models are summarized in Table~\ref{tab:models}.

\begin{table}
	\center
	\caption{List of Models}
	\begin{tabular}{l c c c c c c}
	\hline \hline
	Model				&	Type	&	Simulation	&	Side	&	LF 	&	Matching	&	$\sigma$\\
						&		&				&	(Mpc/h) &	constraint		&			&	(dex) \\
	\hline
	SAMv1	&	SAM		&	Consuelo	&	420	&	K-band	&	-	&	-	\\
	SAMv2	&	SAM		&	Consuelo	&	420	&	None	&	-	&	-	\\
	SHAM	&	SHAM	& 	Consuelo	&	420	&	r-band	&	$\vpeak$	&	0.2	\\
	vp0		&	SHAM	&	Bolshoi	&	250	&	r-band	&	$\vpeak$	&	0	\\
	vp1		&	SHAM	&	Bolshoi	&	250	&	r-band	&	$\vpeak$	&	0.1	\\
	vp2		&	SHAM	&	Bolshoi	&	250	&	r-band	&	$\vpeak$	&	0.2	\\
	vp3		&	SHAM	&	Bolshoi	&	250	&	r-band	&	$\vpeak$	&	0.3	\\
	vn2		&	SHAM	&	Bolshoi	&	250	&	r-band	&	$\vmax$	&	0.2	\\
	Esm		&	SHAM	&	Esmeralda&	640	&	r-band	&	$\vpeak$	&	0.2	\\
	MD		&	SHAM	&	MultiDark	&	1000&	r-band	&	$\vpeak$	&	0.2	\\
	\hline
	\end{tabular}
	\label{tab:models}
\end{table}

These models based on the Consuelo dark matter halos include large variation in the luminosity function.  This variation, as well as the associated differences in M/N and the clustering, demonstrates how different the galaxy formation is in the three cases, and permits a test of whether our method can successfully extract the cosmological parameters from very different models.

We also use the Bolshoi simulation \citep{KTP2011}, which has cosmology $\om=0.27$ and $\sigate=0.82$.  This simulation has the advantage of probing lower mass subhalos, with the drawback of lower volume relative to Consuelo, ~(250~$\mpc/h)^3$ rather than (420~$\mpc/h)^3$.  We will compare a set of models drawn from those presented in \cite{Red2013}, to demonstrate a smoother variation between HODs than the few extreme cases in Consuelo.  The cases we consider will be those matched to the peak maximum circular velocity (or $\vpeak$) of the (sub)halos, with scatter ranging from 0 to 0.3 dex in luminosity at fixed $\vpeak$, and one model using the maximum circular velocity at the present time ($\vnow$) with 0.2 dex scatter.  This smoother variation in the galaxy model allows a more precise test for systematic variation in the inferred cosmology with galaxy formation.

Our final pair of models use two larger boxes.  The Esmeralda simulation, also drawn from the set of LasDamas simulations, has the same cosmology as Consuelo ($\om=0.25$ and $\sigate=0.80$) but larger volume, ~(640~$\mpc/h)^3$. The MultiDark simulation \citep{Rie2011}, has larger volume than Bolshoi, ~(1000~$\mpc/h)^3$, but the same cosmology, $\om=0.27$ and $\sigate=0.82$.  We run the same SHAM model on both simulations, matching to $\vpeak$ with 0.2 dex of scatter.  Since both of these simulations with larger volume have poorer mass resolution, we will consider galaxy populations at lower number density (brighter luminosities) where the missing low-mass halos will not have a large impact.  We will use these two models as a test of the distinction between cosmologies and to demonstrate changes in the constraints with larger volume.

For all the Bolshoi and Consuelo models, we use a luminosity threshold of $M_r <-20.5$ to determine our galaxy sample.  For the Esmeralda and MultiDark simulations, we use a fixed number density cuts of $1.21\times10^{-3}~(\mpc/h)^{-3}$ and $3.00\times10^{-4}~(\mpc/h)^{-3}$ respectively, which correspond to luminosity thresholds of $M_r<-21.0$ and $M_r<-21.447$.

Because the halo model we use is an approximation, there will be physical effects both in the real universe and our simulations that are not incorporated into the HOD.  Using the simulations to make mock galaxy catalogs provides a way to test that the approximations in the HOD do not reduce the efficacy of the HOD approach.  These possible physics effects include:

\begin{itemize}
\item Scatter in the concentration-mass relationship of dark matter halos.  Estimates of this scatter 
in the literature range from $\sim0.1-0.18$ \citep{Bul2001, Wu2013a} for cluster-sized halos.
If satellite galaxies follow the same scatter in concentrations as the dark matter, then the distribution of very small-scale pairs may be different from the no-scatter case.
\item Non-spherical halos.  Any non-spherical distribution of halo substructure, and thus of subhalos and galaxies, could potentially alter the two-point correlation.  Especially on small scales, additional correlations in shape such as between neighboring halos or within halos can enhance the two-point clustering relative to the case with spherical halos.  This is demonstrated by some current studies (e.g., \citealt{vDa2012, Wu2013} and references therein).
\item Deviations of the subhalo distribution from an NFW profile (e.g., \citealt{Gao2008}) or from the dark matter concentration-mass relation.  As discussed in the previous point, systematic deviations of the satellite galaxy distribution in halos from what is expected will generally produce systematic differences in the clustering predicted from a given HOD, particularly at the smallest scales. 
\item Non-poisson scatter in the subhalo or satellite galaxy distribution.  The distribution for the number of satellites in halos of a given mass is generally described as a Poisson distribution with mean $\langle N_{sat}(M_{vir})\rangle$.  While previous works (\citealt{ZBW2005,YMB2008}) have found results consistent with a Poisson distribution, there exists a possibility of deviations from Poisson (e.g., \citealt{Kra2004}), particularly when the mean number of satellites is small.  In turn, a difference in this distribution of satellites at low host halo mass or richness has the potential to alter the two-point clustering at small or moderate scales, near the transition between the one-halo and two-halo terms.
\item Assembly bias.  As discussed in, e.g., \cite{GSW2005, CGW2007}, assembly bias alters the distribution of galaxies, even in the case where the mean number of galaxies per halo is held constant.  That is, the number of galaxies per halo depends on both the mass of the halo and how it was formed (such as earlier or later formation).  In this case, for a given HOD, the M/N measurement would remain the same, but the two-point clustering would differ from that which would be predicted without assembly bias.  The extra dependence on formation time or another parameter could potentially biased constraint.
\end{itemize}

In short, there are a number of approximations in our HOD framework.  Thus, in addition to testing different models of galaxy formation, it is necessary to demonstrate that none of these approximations has a deleterious effect on our ability to recover cosmological parameters in an unbiased fashion.  In the following section, we will show that we recover the true cosmology in all cases, even when tested with Gpc-sized simulations that yield clustering measurements with errors on the order of a few percent.

\begin{figure}
\includegraphics[width=0.5\textwidth]{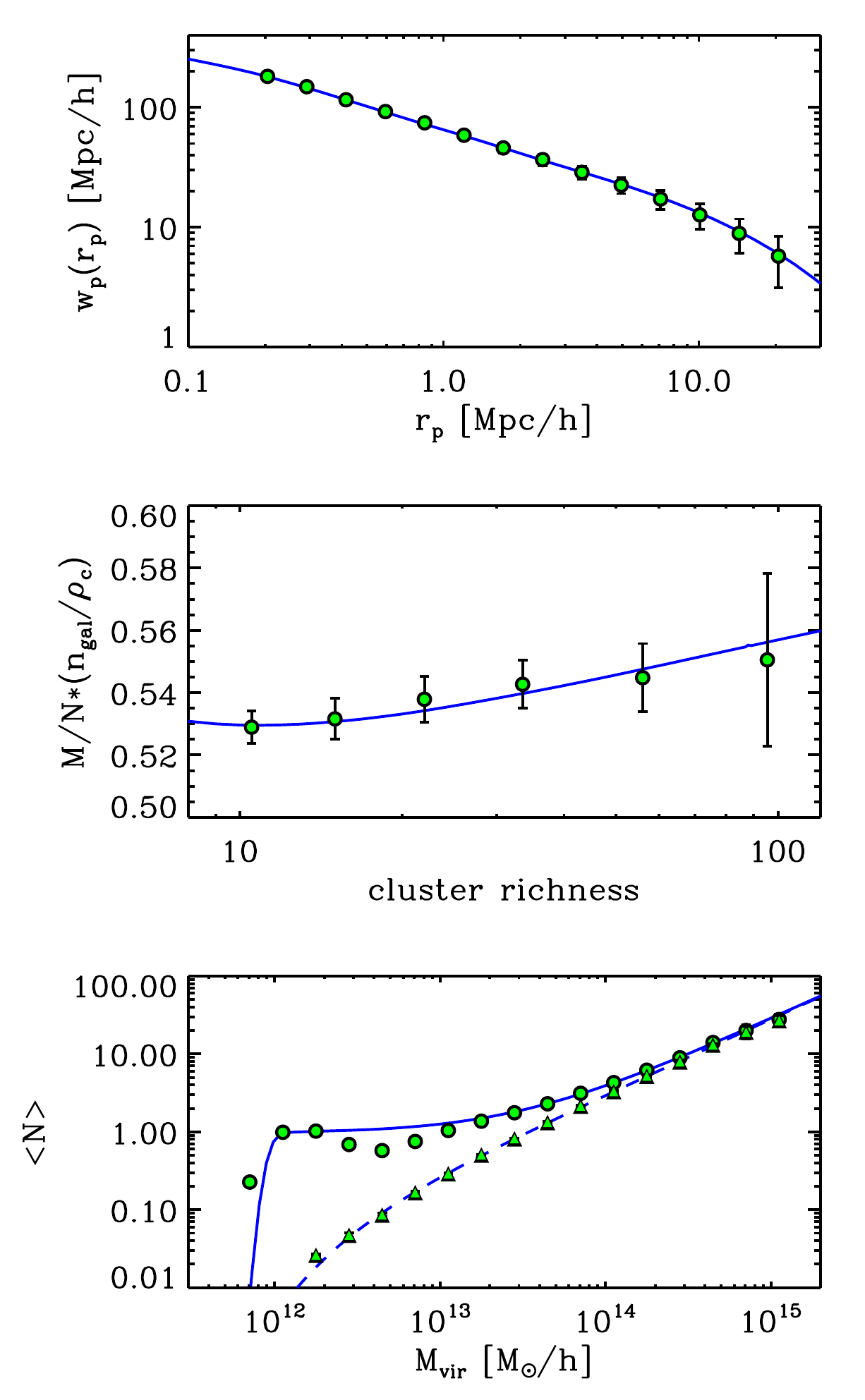}
\caption{Demonstration of fit to $w_p(r_p)$, HOD, and the mass-to-number ratio M/N.  Points are the measurements from the galaxy model, while the blue lines show the best fit to the $w_p(r_p)$ and M/N.  The galaxy formation model shown is SAMv1.  {\em Top:}  Projected two-point clustering.  {\em Center:} M/N ratio.  These are the two measurements our method uses as input.  {\em Bottom:} HOD.  Circles indicate the total HOD, and triangles the satellites only, in the galaxy model.  Solid line is overall HOD in the best-fit case, and the dashed line is the satellites only.  Note that the method we discuss does not fit the HOD directly.}
\label{fig:fit_gen}
\end{figure}

\section{Methods}\label{sec:method}

We combine two measurements to reconstruct the cosmology.  The first is a projected two-point correlation function (e.g., \citealt{Zeh2005, Zeh2011, WBZ2011, Nuza2012} and references therein).  We obtain errors by scaling to the variation in clustering in multiple other simulation boxes of the same resolution (for Consuelo) or in a set of PM boxes of adequate resolution for the scales of interest (Bolshoi).  In all cases, we use the full covariance matrices.

\begin{figure*}
\includegraphics[width=0.95\textwidth]{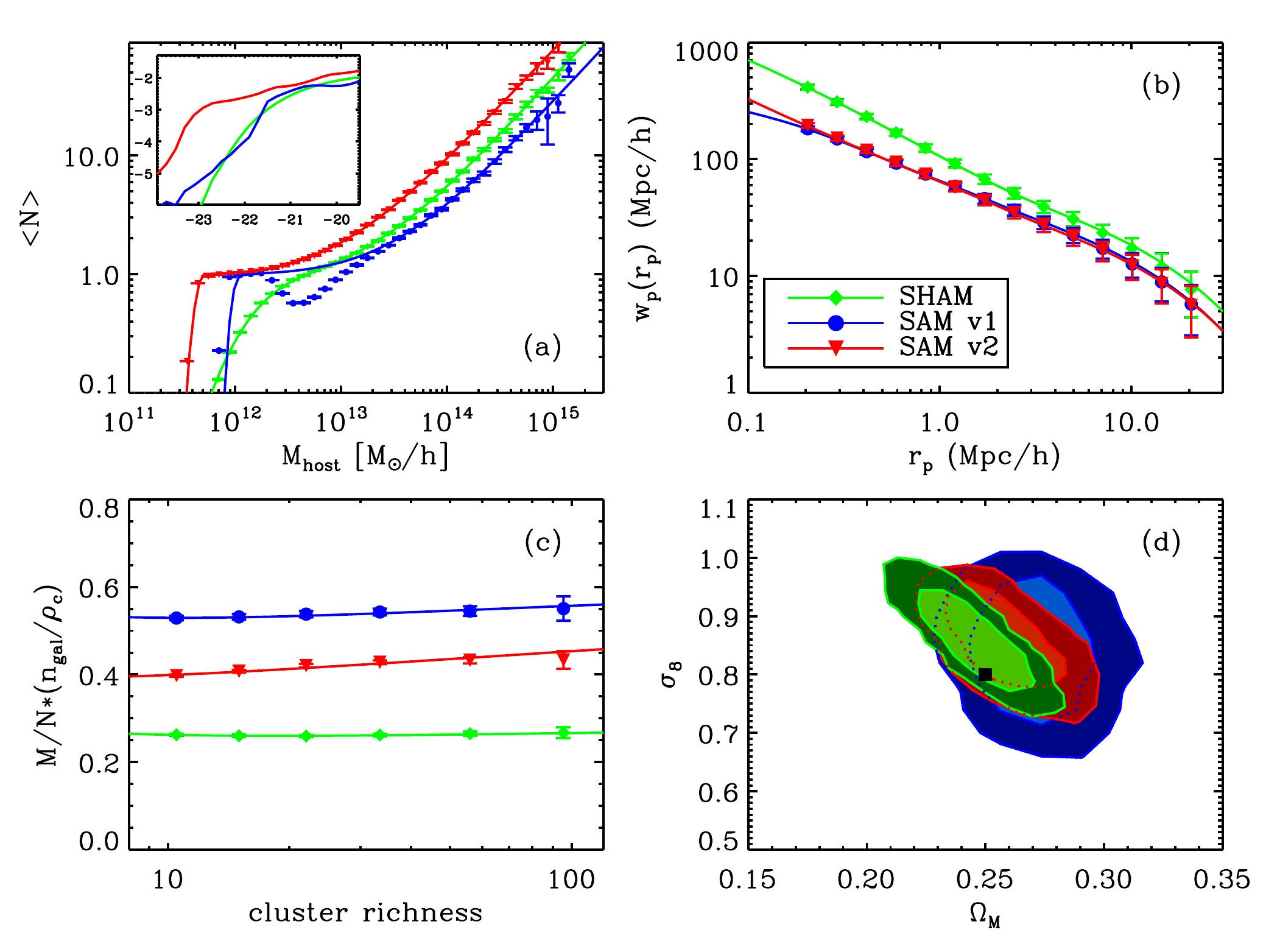}
\caption{ Applying the HOD method to three
  different galaxy formation models built upon the same N-body
  simulation (Consuelo).  Panel (a): Luminosity functions. SAMv1 and SAMv2
  represent two semi-analytic galaxy formation models. SHAM used the
  abundance matching method.  {\em Inset:}  Luminosity functions used in each model.  Axes are r-band magnitude and base-10 log of $\Phi$ $[(Mpc/h)^{-3}~\rm{mag}^{-1}]$.  Panel (b):
  Points with error bars show the projected clustering measurements
  from each model. Curves show best-fit HOD models, where both halo
  occupation and cosmological parameters are free. Panel (c): $M/N$
  measurements, as a function of cluster richness, from the galaxy
  formation models. Panel (d): Constraints on cosmological parameters
  after marginalizing over HOD parameters. The true cosmology is
  indicated with the solid square.}
\label{fig:cosmo_cons}
\end{figure*}

The second measurement is the M/N value, which we obtain by first assigning a "richness" to the isolated halos.  This is to emulate observations where an optical richness is used as a mass proxy.  We assume a mass-richness relationship of the form:

\be\label{eq:n200}
	\bar{N}_{200} = 1.09+0.75\cdot\ln(\mvir/M_{\rm piv})
\ee

We use a value of $M_{\rm piv}=1.44\cdot10^{13} \Msun/h$, and include a Gaussian scatter of 0.35 in $\ln N$ (or 0.15 dex) in richness at fixed mass.  This form is that of \cite{Rozo2009} changed to $h^{-1}$ units.

To recover the expected M/N ratio for each bin in richness $N_{200}$, we evaluate an expression similar to that presented in \cite{Tin2012}:

\be
	M/N = \frac{\Sigma_{\nrich} \int dM n_h(M) P(\nrich|M) M}{\Sigma_{\nrich} \int dM n_h(M) P(\nrich|M) \langle N_{\rm{sat}} \rangle_M}
\ee

Here, $n_h(M)$ is the halo mass function.  $P(\nrich|M)$ is the distribution of richnesses at fixed mass, a log-normal as described above.  $\langle N_{\rm{sat}} \rangle_M$ is the mean number of satellites above the luminosity threshold in a halo of mass M.  Covariance matrices are estimated via jackknife resampling of the galaxy simulation.

We simultaneously fit the HOD and the cosmological parameters $\om$ and $\sigate$ using an HOD model combined with the halo mass function of \cite{Tin2008} and bias functions of \cite{Tin2010}.  We fit the two-point clustering and the M/N measurements using a six-parameter HOD of the form:

\be
	\langle N \rangle = \langle N_{\rm{cen}}\rangle + \langle N_{\rm{sat}} \rangle 
\ee
\begin{align}
	\langle N_{\rm{cen}} \rangle & =  \frac{1}{2} \left[1+\mathrm{erf} \left( \frac{\log \mvir-\log M_{\rm{min}}}{\sigma_{\log M}}\right)\right] \nonumber \\
						       &\times {} [1+0.05\log(\mvir/M_{lin}) ]
\end{align}
\be
	\langle N_{\rm{sat}} \rangle = \left( \frac{\mvir}{M_1}\right)^{\alpha_{\rm HOD}} \exp\left(-\frac{M_{\rm{cut}}}{\mvir}\right)
\ee

\begin{figure*}[t!]
\includegraphics[width=0.95\textwidth]{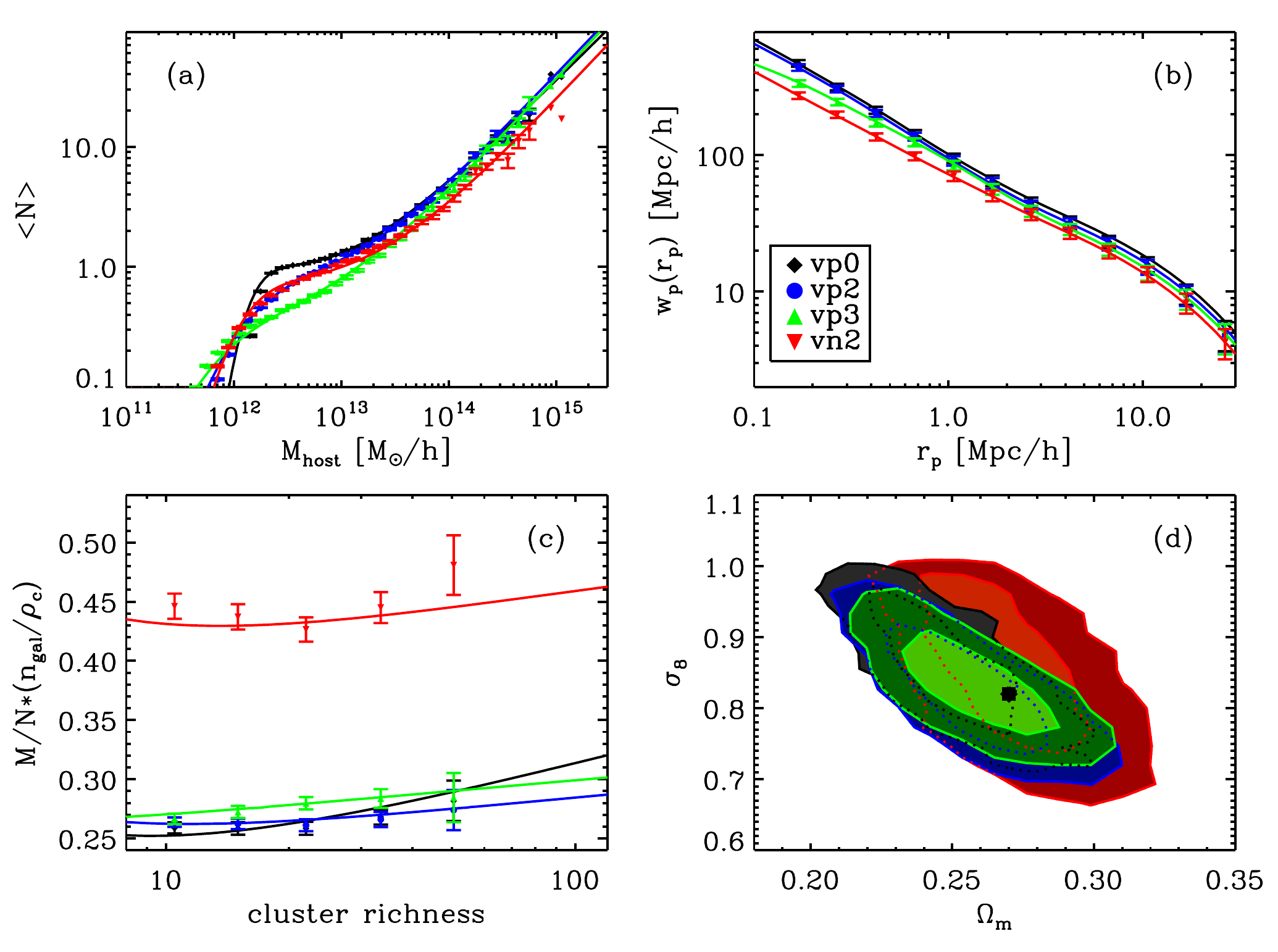}
\caption{Similar to Fig.~\ref{fig:cosmo_cons}, we apply the HOD method to five different abundance matching models on the same Bolshoi simulation.  Because these models use abundance matching, the luminosity functions are identical.  The "vp" models use $\vpeak$, while "vn" indicates $\vnow$.  The number indicates the scatter in tenths of dex.  The $\sigma=0.1$ model is very similar to the zero-scatter model, and has been omitted for the sake of clarity.  Panel (a): HODs for the different models. Panel (b):
  Points with error bars show the projected clustering measurements
  from each model. Curves show best-fit HOD models, where both halo
  occupation and cosmological parameters are free. Panel (c): $M/N$
  measurements, as a function of cluster richness, from the galaxy
  formation models. Panel (d): Constraints on cosmological parameters
  after marginalizing over HOD parameters. The true cosmology is
  indicated with the solid square.}
\label{fig:cosmo_bols}
\end{figure*}

We find that the recovery of the cosmological parameters is insensitive to small differences in the analytic form of the HOD, so long as the HOD model is capable of capturing the important features.  As will be demonstrated in the SAM models, differences in the HOD at the low-mass end may be compensated for.  However, unusual HODs at the massive end can complicate our procedure.  This is especially problematic for HODs where the fraction of halos with a central galaxy brighter than the chosen threshold is significantly less than one.  In the models we have considered, this occurs for thresholds with roughly $M_r<-21.5$.  We discuss the necessity of using a more complex form for the HOD in Appendix~\ref{app-hod}.

To calculate the clustering from the HOD, galaxies are assumed to follow an NFW profile \citep{NFW1997,Guo2012,Tin2012}, with a concentration of $f_{con}\times c_{\rm{vir}}$, where $c_{\rm{vir}}$ is the host halo concentration.  The M/N ratio is calculated using the HOD and the $\bar{N}_{200}$ relationship given in Eq.~\ref{eq:n200}.  The concentrations are based on the model of \cite{Bul2001} using the updated parameters from \cite{WZB2006}.  A test using the concentration-mass relation of \cite{Pra2012} did not show any significant difference in the estimation of parameters aside from $f_{con}$.  Therefore, we do not expect that the precise details of the concentration-mass relation will significantly affect our results.

Because the value of $M_{\rm{min}}$ is set by the number density of galaxies, we are left with five free HOD parameters, and two free cosmological parameters ($\om$ and $\sigate$).  We then use a Markov Chain Monte Carlo method to explore this parameter space with flat priors.

As described in \cite{Tin2012}, fitting to the clustering alone results in degeneracies between the HOD parameters and the cosmology.  Adding the M/N measurement breaks this degeneracy by placing a strong constraint on the power law part of the HOD.  This restricts the possible number of satellite galaxies, and therefore the small-scale clustering in particular.  An example fit to SAM v1 is shown in Fig.~\ref{fig:fit_gen}.  The $w_p(r_p)$ and M/N measurements on the SAM are clearly well-reproduced.

There are some degeneracies that remain in the model.  The shape of the cutoff of the HOD at low host halo masses for central galaxies is not always well reproduced, as is clear in Fig.~\ref{fig:fit_gen}.  In this regime, the M/N ratio provides no constraint, as halos there are too low mass and have too few galaxies for M/N to be reliably measured.  Thus, there is a significant degeneracy between the parameters that control the cutoff, which is constrained only by the two-point clustering in the transition region between the 1-halo and 2-halo terms.  Nonetheless, we find that fits to the models are more accurate if all three cutoff parameters ($M_{\rm{min}}$,  $\sigma$, $M_{lin}$) are included.  In the particular case of SAM v1, there is an extra feature in the HOD produced by quenching that is not captured by the form of the HOD we have chosen.  A good fit to the M/N and clustering does exist, though the precision of the measured cosmological parameters is reduced.  While we do not expect such an unusual HOD to occur physically, it is encouraging that the fit still recovers reasonable cosmological results.

For the large-volume models (using the Esmeralda and MultiDark simulations), prior to running the MCMC chain, we constrain the scale-dependent halo bias.  For the other models, we use the parameterization and exact parameters presented in \cite{Tin2005}.  That parameterization is given by:
\be
	b^2(M,r)=b^2(M)\frac{[1+1.17\xi_m(r)]^{1.49}}{[1+0.69\xi_m(r)]^{2.09}}
\ee
where $\xi_m(r)$ is the non-linear matter clustering.  For all the smaller volume simulations, the numerical factors in this equation are held fixed, which is sufficient for the true HOD measured from the simulation to accurately reproduce the two-point clustering.  However, this is not the case for Esmeralda and MultiDark.  Therefore, we separately fit these numerical factors in those cases, while holding the HOD and cosmology fixed to the truth values.  This is necessary to ensure that the modeling of the $w_p(r_p)$ from the true HOD is accurate.  The previous study in \cite{Tin2012} stated that $b(r)$ was one of the main systematic uncertainties in the analysis.  Our investigation demonstrates that b(r) is not mass-independent. In a future work, we will present a new calibration of scale dependent bias that fully takes into account the dependence on halo mass.

\begin{figure*}
\includegraphics[width=0.95\textwidth]{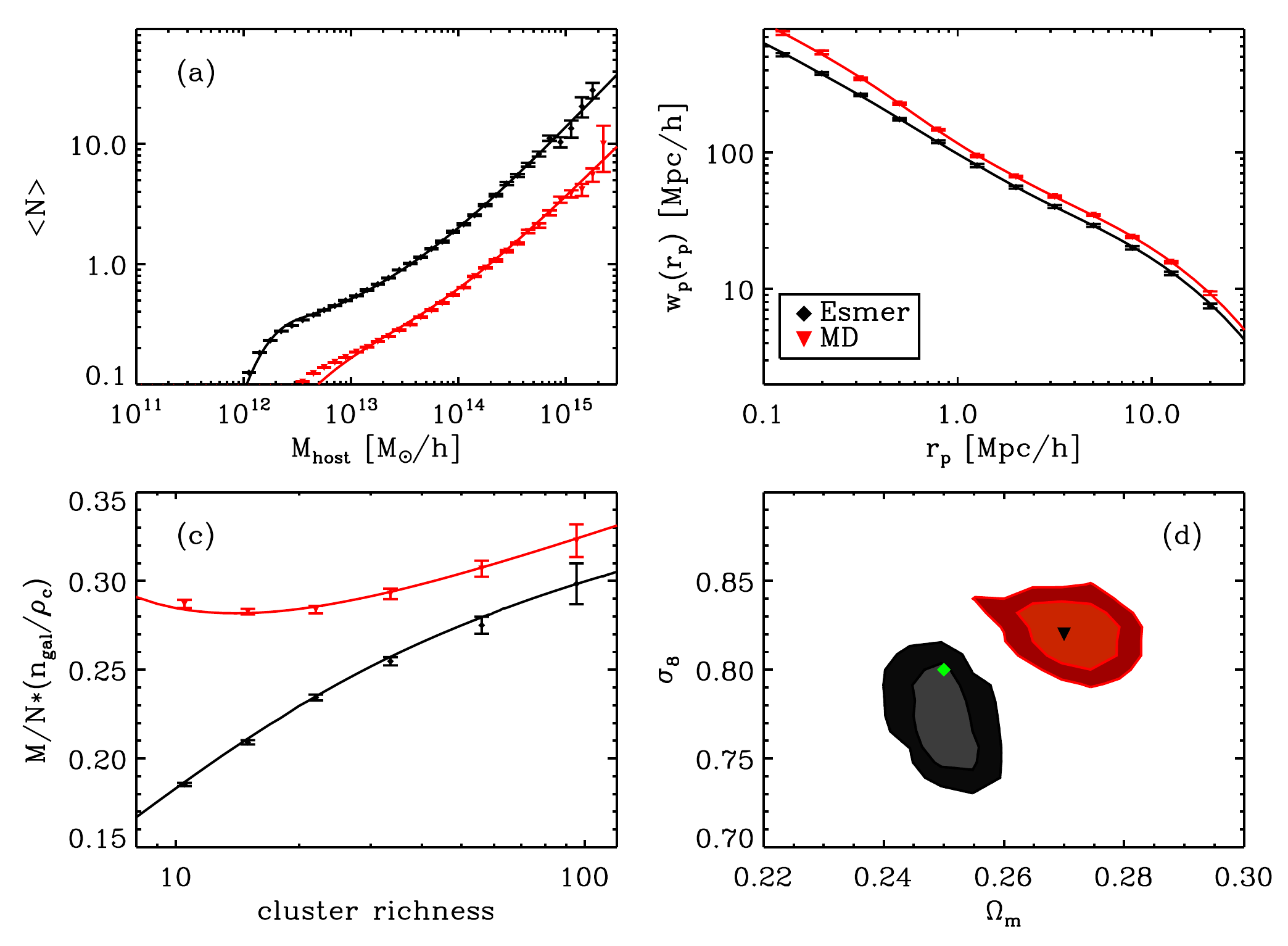}
\caption{Similar to Fig.~\ref{fig:cosmo_bols}, we apply the HOD method to two different abundance matching models on two different large-volume simulations.
  Panel (a): HODs for the different models. Panel (b):
  Points with error bars show the projected clustering measurements
  from each model. Curves show best-fit HOD models, where both halo
  occupation and cosmological parameters are free. Panel (c): $M/N$
  measurements, as a function of cluster richness, from the galaxy
  formation models. Panel (d): Constraints on cosmological parameters
  after marginalizing over HOD parameters. The true cosmology for the Esmeralda-based model is shown with a green diamond; the MultiDark cosmology is shown
  with a black triangle.}
\label{fig:cosmo_es}
\end{figure*}

\section{Results}
\label{sec:results}

The results using the Consuelo models are shown in Fig.~\ref{fig:cosmo_cons}.  Fig.~\ref{fig:cosmo_cons}a shows the very different luminosity functions of each model, as well as the different clustering and M/N results for the $M_r<-20.5$ sample.  However, all three galaxy models are built on the same halo population.  Thus, the differences in clustering and M/N are due only to differences in the halo occupation.  The models are well-separated in halo occupation, as is shown by the M/N ratio.  The constraints in the $\om-\sigate$ in Fig.~\ref{fig:cosmo_cons}d show the result of applying our method, marginalizing over the HOD.  In all three cases, the results are consistent with the true cosmology.  The larger area of the SAM v1 contours is largely due to the difficulty in accurately reproducing the non-monotonic HOD that is present in that model.

The Bolshoi models are shown in Fig.~\ref{fig:cosmo_bols}, in the same format as given in Fig.~\ref{fig:cosmo_cons}.  We omit the model with scatter of 0.1 dex from this plot for the sake of clarity due to its similarity to the zero-scatter case.  Because the luminosity function is the same for all of the SHAM galaxy models used on the Bolshoi simulation, we instead show the HODs in Fig.~\ref{fig:cosmo_bols}a.  In all cases, the contours in Fig.~\ref{fig:cosmo_bols} are consistent with the input cosmology.  For this family of models, we also find that the contours are very similar to each other, regardless of the quantity of scatter.  This is not surprising, as the HOD for these models is very similar for massive halos, with differences only becoming clear below the halo mass where M/N is measured.  The significantly reduced number of galaxies in the $\vmax$ model increases the variance in the M/N measurement, somewhat increasing the area of the contour and altering its shape.

The models using significantly larger volumes in the Esmeralda and MultiDark simulations are shown in Fig.~\ref{fig:cosmo_es}.  We also note here that rather than using the M/N ratio with N as the number of satellites, the analysis on Esmeralda uses the ratio between the mass and the total number of galaxies, M/N$_{tot}$.  This helps avoid some of the complications with deviations of the central HOD from our fitting function, which occurs at relatively high masses for sufficiently bright luminosity thresholds.  Ultimately, we expect that using all of a cluster's galaxies when working with observations will help avoid additional systematic errors introduced by miscentering (e.g., \citealt{RoRy2013}).  Regardless, analyses on both simulations provide relatively tight and accurate cosmological measurements, either within the 1-$\sigma$ contours (MultiDark) or just on the edge of 1-$\sigma$ (Esmeralda).  Note the smaller axes in Fig.~\ref{fig:cosmo_es}(d) relative to panel (d) in the previous figures.

\begin{figure}
\includegraphics[width=0.5\textwidth]{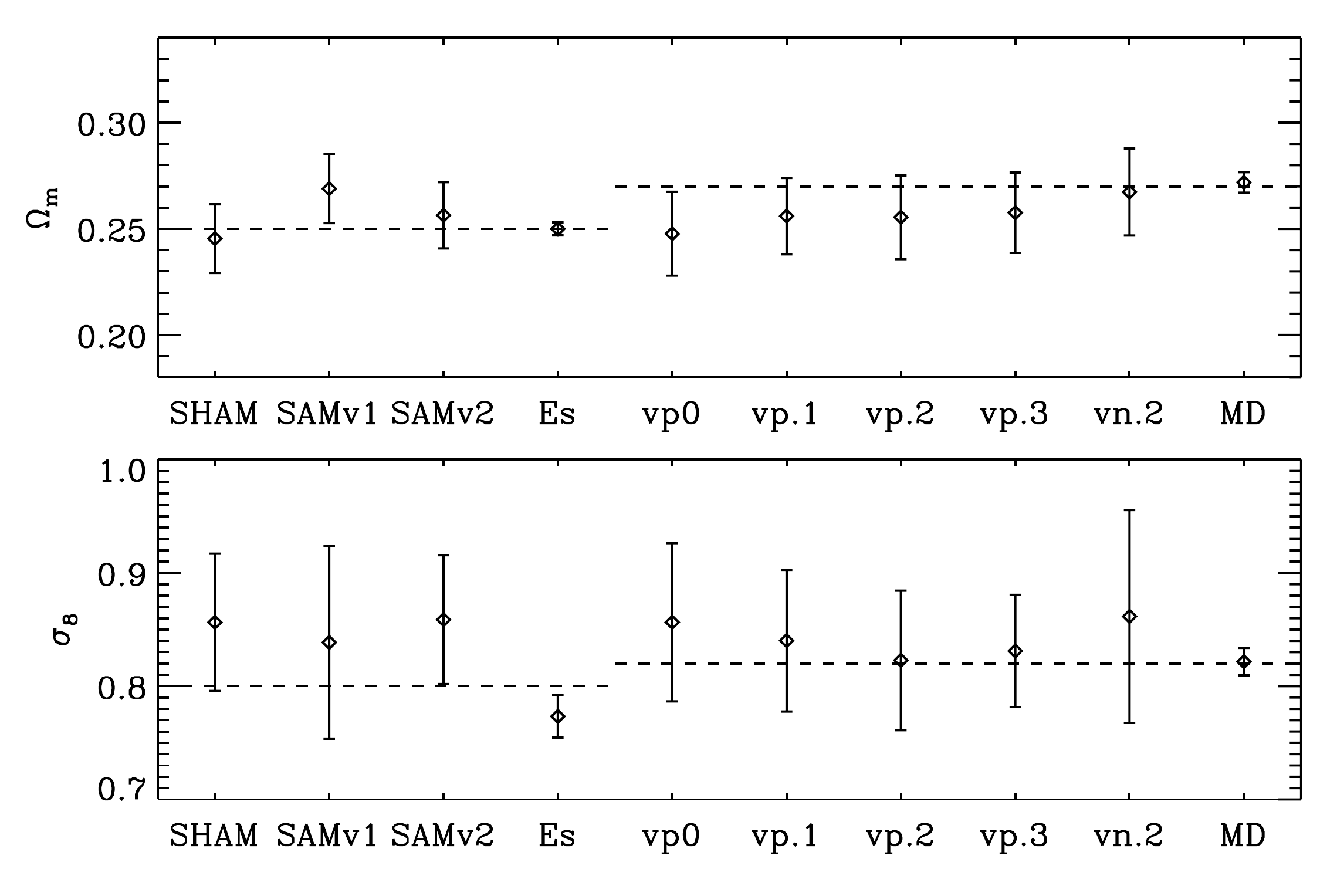}
\caption{Values of $\om$ and $\sigate$ recovered from our analysis, compared against the input.  Diamonds indicate the median result, and error bars give the 68\% bounds.  Dashed lines indicate the true value from the relevant simulations.}
\label{fig:cosmo_all}
\end{figure}

To summarize all of our models, Fig~\ref{fig:cosmo_all} shows the best fit results and the marginalized errors for all of the models.  While in some cases the truth is outside the marginalized 68\% confidence limits, the results are consistent in the $\om-\sigate$ plane.

\section{Summary and Conclusions}\label{sec:conclude}

We investigate the combination of galaxy clustering and the mass-to-number ratio of galaxies in clusters as a cosmological probe, first introduced by \cite{Tin2012}.  Here, we focus on the ability of this method to obtain robust cosmological constraints while marginalizing over galaxy formation.  We consider four cosmological boxes, with different galaxy formation prescriptions, number densities, clustering properties, luminosity functions, underlying halo occupations, and box volumes. For all ten models we examine, we are able to recover the correct $\om$ and $\sigate$ values within the 68\% contour.  This suggests that for data samples of this size, we are effectively able to marginalize
over the physics of galaxy formation and its consequent galaxy--halo connection to obtain unbiased cosmological 
parameters.  We suggest that this kind of analysis should become standard procedure for 
all cosmological constraints that depend on galaxy properties or the galaxy--halo connection.

Of some interest is the dependence of our results on the analytic form of the HOD we choose.  Our current model is clearly adequate for the galaxy models we present here.  However, a more extreme HOD than that of SAM v1, with a much larger decrement in the HOD, was only poorly fit and gave biased cosmological parameters.  We expect that a such non-monotonic HOD is not physical.  However, for non-standard galaxy samples, for example those selected by color or star formation rate rather than by luminosity, care should be taken to assure that the analytic form of the HOD is flexible
enough to model the data.

In future work, we expect to obtain tighter constraints by using multiple different luminosity thresholds on the same data set, or by using multiple redshifts.  We anticipate applying this method to data from the Sloan Digital Sky Survey,
the BOSS survey, and the Dark Energy Survey.  We may apply this method to cluster galaxies only, rather than all galaxies, which may eliminate some of the uncertainties associated with the low host halo mass cutoff in the HOD.  Angular clustering, combined with accurate photometric redshifts, will allow application of this method to the large volumes 
being probed by photometric surveys.  Further work will be required to show to robustness of this method to photometric redshift errors and over wide redshift ranges.

\section*{Acknowledgements}

This work has been supported by the National Science Foundation under NSF-AST-1211838, and by Stanford University, through a Stanford Graduate Fellowship to RMR.  We thank the participants of the Fall 2012 ``Small-Scale Structure'' workshop at Stanford for useful discussions.  We are grateful to Anatoly Klypin for providing access to the Bolshoi and MultiDark simulations, and to our collaborators in the LasDamas collaboration for providing access to the Consuelo and Esmerelda simulations.  We thank Peter Behroozi for providing the halo catalogs and merger trees for all four simulations.
\bibliography{mngal}

\newpage

\begin{appendices}
\section{Parameterization of the HOD}\label{app-hod}

We use a slightly more complex variation on the HOD relative to other works based on the halo occupation distribution (e.g., \citealt{Zeh2011}) because we find that incorrect modeling of the central part of the HOD can cause large deviations in the two-point clustering, particularly for bright thresholds.

\begin{figure}
\centering
\includegraphics[width=0.6\textwidth]{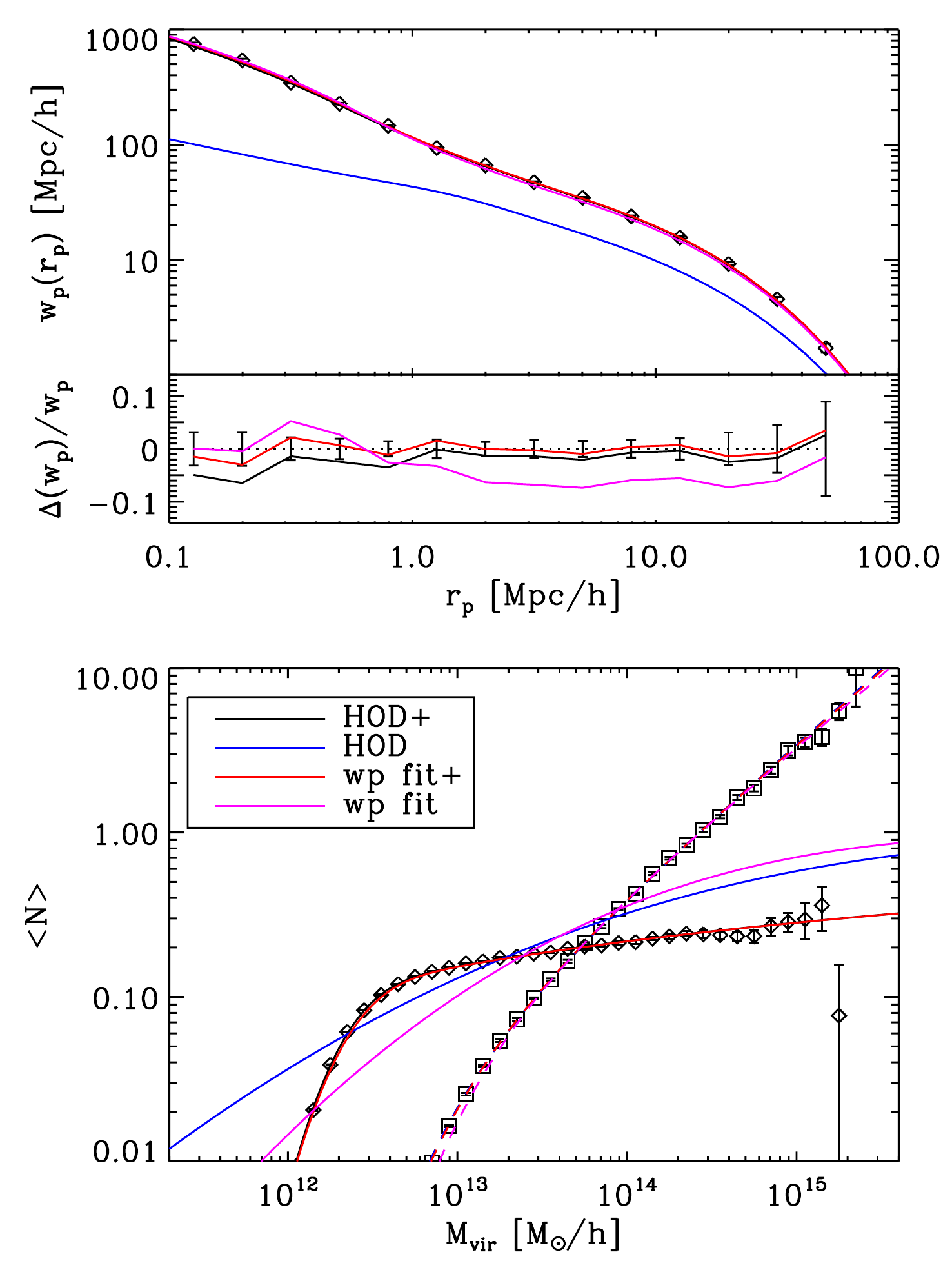}
\caption{Comparison of different HOD models.  {\em Top:}  Black points show the two-point correlation measured on MultiDark.  Lines are the correlation functions calculated from different HODs.  The subplot shows the difference between the values measured in the simulation and the clustering predicted from the HOD models.  {\em Bottom:}  Black diamonds are the measured central HOD, and black squares the measured satellite HOD.  Lines are the HOD models corresponding to the  clustering shown in the upper plot.  Solid lines are the central HOD, and dashed lines the satellite part.  The ``HOD'' models are fit to the HOD measured in the simulation rather than to the $w_p(r_p)$ and M/N, and used to predict the $w_p(r_p)$.  The HOD line uses only a five-parameter model, with two parameters for the central part, while the HOD+ version uses two additional parameters to describe the shape of the central part of the HOD.  The ``$w_p$'' models are fit to the $w_p(r_p)$ and the M/N, with cosmology fixed.  Again, the plus denotes a fit using additional parameters.  The models with additional parameters clearly provide a better fit to the $w_p(r_p)$ in both cases.}
\label{fig:hod_demo}
\end{figure}

An example of this is shown in Fig.~\ref{fig:hod_demo}.  Here, we present a comparison of the true HOD and clustering from the simulation (points) compared against several different instances of the HOD model.  In all these cases, we hold the cosmology fixed.  Restating our HOD model, we have:

\begin{align}
	\langle N_{\rm{cen}} \rangle & =  \frac{1}{2} \left[1+\mathrm{erf} \left( \frac{\log \mvir-\log M_{\rm{min}}}{\sigma_{\log M}}\right)\right] \cdot \left[1+f_{cen}\log\left(\frac{\mvir}{M_{lin}}\right) \right]
\end{align}
\be
	\langle N_{\rm{sat}} \rangle = \left( \frac{\mvir}{M_1}\right)^{\alpha_{\rm HOD}} \exp\left(-\frac{M_{\rm{cut}}}{\mvir}\right)
\ee

This model reduces to the more standard five-parameter model if $f_{cen}$ is set to zero, which also ensures that $M_{lin}$ no longer has any effect.  We demonstrate the difference between these two HOD models as applied to our MultiDark sample in Fig.~\ref{fig:hod_demo}.

Our first comparison is between the HOD and HOD+ models, keeping our cosmology fixed to the true values.  The ``HOD'' models are produced by fitting to the HOD directly, and then using that specific model to predict the clustering.  While the seven-parameter model produces a good fit (HOD+), the other does not.  This is because the only way to reproduce the lack of centrals at high halo mass in the five-parameter model is to increase the $M_{min}$ threshold and greatly increase the width $\sigma_{\log M}$.  The clustering produced in this case is severely suppressed due to the inclusion of many low-mass halos, which also forces the number density to be far too high.

We also test fitting each model to the $w_p(r_p)$ and M/N statistics, allowing only the HOD parameters to vary.  For our best first, we obtain a $\chi^2$ of 19 for our seven-parameter model (13 degrees of freedom).  This is a marginally good fit ($P[\chi^2<19]\approx0.1$).  The model with five parameters has a $\chi^2$ of 119 (15 degrees of freedom), which is clearly not acceptable.  Visually, the latter model is clearly insufficiently clustered on moderate to large scales ($r_p\gsim1~\rm{Mpc/h}$).  Setting the cosmological parameters free in this case would bias $\Omega_m$ low and $\sigma_8$ high relative to the true values.

\end{appendices}

\end{document}